# Molecular information theory meets protein folding

Ignacio E. Sánchez*, Ezequiel A. Galpern, Martín M. Garibaldi, Diego U. Ferreiro*

Facultad de Ciencias Exactas y Naturales, Laboratorio de Fisiología de Proteínas, Consejo Nacional de Investigaciones Científicas y Técnicas, Instituto de Química Biológica de la Facultad de Ciencias Exactas y Naturales (IQUIBICEN), Universidad de Buenos Aires, Buenos Aires, Argentina.

* Correspondence may be addressed to IES (isanchez@qb.fcen.uba.ar) and DUF (ferreiro@qb.fcen.uba.ar). Tel +54-11-45763300 ext 205

**Abstract**

We propose an application of molecular information theory to analyze the folding of single domain proteins. We analyze results from various areas of protein science, such as sequence-based potentials, reduced amino acid alphabets, backbone configurational entropy, secondary structure content, residue burial layers, and mutational studies of protein stability changes. We found that the average information contained in the sequences of evolved proteins is very close to the average information needed to specify a fold ~2.2 ± 0.3 bits/(site operation). The effective alphabet size in evolved proteins equals the effective number of conformations of a residue in the compact unfolded state at around 5. We calculated an energy-to-information conversion efficiency upon folding of around 50%, lower than the theoretical limit of 70%, but much higher than human built macroscopic machines. We propose a simple mapping between molecular information theory and energy landscape theory and explore the connections between sequence evolution, configurational entropy and the energetics of protein folding.

**Keywords**

Protein folding, information theory, configurational entropy, energy landscape theory, selection temperature, reduced alphabet

## 1. Introduction

Scientific research can be said to advance one metaphor at a time (1). Good metaphors allow us to think with soft-concepts (2), generate challenging hypotheses, translate between disciplines and eventually communicate complex findings in pedagogical terms. Yet philosophers warn us to be aware of these double edge swords as metaphors can also constrain scientific reasoning, contribute to public misunderstandings and inadvertently reinforce stereotypes that undermine the goals of science (3). Metaphors in biological research are so ubiquitous that we have to some extent become blind to their existence. Some of the most used (and misused) metaphors in contemporary Biology deal with the concept of Information. On the many aspects of the influence of Information in biological sciences (4), we will concentrate here on the most basic, molecular, application.

Biomolecules have physicochemical characteristics that distinguish them from other large organic molecules: collectively they self-reproduce, they fold into amazingly precise ensembles of structures, they catalyze reactions, they respond to apparently gratuitous environmental changes, they appear to be 'informed' molecules that 'know how to do things' in the most teleonomic (5) or robotic (6) sense of the words. It is usually believed that these characteristics have been somehow 'coded' in their structures by evolutionary processes. As some scholars put it in Lamarckian overtones, evolution is a process where information flows from the environment into the genomes (7, 8). One of the most powerful uses of the information metaphor appears in the study of protein folding and function.

Natural terrestrial proteins are composed by linear polymers of just a few types of alpha-amino acids that, in the appropriate environmental conditions and short time-scales, typically fold and perform chemical activities that relate to biological functions (9). On longer time-scales the sequences of proteins change and this may affect the populations of the structural ensembles and their dynamics, impacting their biological role. Today's implicit metaphor is that the sequence of amino acids somehow encodes the biological needs of the protein (10, 11), and thus the polymer somehow computes from this input its own structure and dynamics as output (12). Yet, the sequences of most natural foldable proteins sometimes appear to be as complex as random strings of amino acids (13, 14), but the polymers built from random amino acids strings are by the most part non-foldable (10). So where is the information in proteins after all? This apparent paradox may be related to the imprecise use of the informational metaphor, which we will try to clarify in this work.

The current paradigm to understand protein folding and its connection to protein evolution is the Energy Landscape Theory, to which our celebrated José Onuchic made pioneering contributions (too many to cite here! please refer to the first pages of this Festschrift issue). One of the most seminal hypotheses he put forward is that the folding landscape of evolved proteins resembles a rough funnel (15). This was unfortunately depicted as 'just a metaphor' by those who didn't grasp it completely, but the evidence for it is now, again, too large to cite here (16–18). Funneled energy landscapes are navigated by systems that follow the Principle of Minimal Frustration (19), for which there is a fundamental correlation between similarity in structure and energetic distribution. The funneled criterion has been postulated to occur both in structure and in sequence spaces (20, 21). If random unevolved polymers have rugged landscapes and are unfoldable frustrated systems, the structural-energetic correlations in folded proteins must somehow account for the 'folding codes' (22, 23). The evolutionary imprint on the sequences of existing protein structural families may contain enough information to reverse-engineer the folding codes, but these do not appear to be simple enough to print on a T-shirt. Recently, machine learning has somehow captured some aspects of these codes (24), and some of us would still like to understand them too (25). Taking the informational metaphor one step further, risking to take the sword by the wrong end, we may ask: what are the characteristics of the 'machine' that 'reads' and 'processes' this 'coded' biological 'information'? For this we will mostly rely on the Molecular Information Theory developed by Thomas D. Schneider (26–43) originally developed for protein-DNA interactions and further applied to other types of molecular machines (41). Here we will extend it to explore how it can be applied to protein sequences, structures, energetics and evolution.

## 2. Results and discussion

### 2.1. Introducing molecular information theory to protein folding

In molecular information theory, a molecular machine is any molecule or macromolecular complex that performs an isothermal operation of specific molecular recognition in a living system, such as proteins binding specific DNA sequences or detecting photons of a specific wavelength (28, 29, 41, 42). This operation starts with priming of the machine to a high-energy state. After priming, any excess energy is quickly dissipated. This leaves the molecule trapped in a flexible "before" macrostate at the ambient temperature in which the machine randomly searches through various conformations and finds the correct one for the operation (28, 29, 41, 42). This transition involves choice amongst multiple possible "after" macrostates. The resulting reduction in uncertainty $H$ upon going from the "before" to the "after" state can be understood as a gain in information $R$ and measured in bits per molecular machine operation by applying Shannon's information theory (34, 38, 44):

$$R(bits/operation) = H_{before} - H_{after} \qquad (1)$$

In this work, we find it convenient to define the uncertainty $H$ and the information $R$ in bits per protein site and operation. Following Shannon's definition,

$$H(bits/(site \cdot operation)) = - \sum_i p_i log_2(p_i) \qquad (2)$$

where $p_i$ is the frequency of amino acid $i$ at the position.

Protein folding can be understood as the self-recognition operation of a molecular machine, in which a specific set of structures is selected from all conformations allowed by polypeptide covalent chemistry. Figure 1A summarizes the operation cycle of a folding molecular machine, starting with a resting folded state that is primed by an energy source, such as thermal noise or energy-dependent processes such as protein synthesis or chaperones. Priming leads to a "before" state in which the protein is unfolded, i.e., does not adopt a specific conformation. The folding machine then relaxes into one of the many possible "after" macrostates, a well-defined conformation called native state, dissipating energy and gaining information in the process.

The reduction in the uncertainty about the protein upon going from the unfolded "before" state to the native "after" state can be quantified from multiple viewpoints. For example, we can measure the information required to find the native conformation in the pool of all polypeptide conformations (26). We propose $R_{Levinthal}$ as a name for this quantity, since it is a quantification of the entropic challenge in Levinthal's paradox (45). We define $R_{Levinthal}$ as the information gained per site and operation upon going from an effective number of configurations of a residue in the unfolded state, $N_{unfolded}$, to a single configuration for each residue in the folded state ($N_{native}$=1).

$$R_{Levinthal}(bits/(site \cdot operation)) = H_{unfolded} - H_{native} = log_2(N_{unfolded}) - log_2(N_{native}) \quad (3)$$

Only a fraction of all possible sequences of a given length are compatible with a given protein fold (46). As a consequence, protein folding is a major evolutionary pressure for many proteins and leaves an information footprint in their sequences. Molecular information theory quantifies this footprint as $R_{sequence}$, the decrease in sequence uncertainty between a protein of random sequence and the ensemble of sequences evolved to be compatible with a given fold (26). If we consider the information gained per site and operation:

$$R_{sequence}(bits/(site \cdot operation)) = H_{random} - H_{evolved} = log_2(N_{random}) - log_2(N_{evolved}) \quad (4)$$

This is often analyzed for protein sequence alignments with gaps, where $N_{random}$ is 21 if we consider all amino acids that are allowed at a given site equiprobable. In this case, $R_{sequence}$ for a sequence site can range from 0 (if the evolved and random states present the same effective alphabet size) to $log_2(21) \approx 4,39$ bits/(site operation). A major prediction of molecular information theory is that the information footprint of the self-recognition process on protein sequences, $R_{sequence}$, should be the same as the information required to find the folded state among all possible chain configurations, $R_{Levinthal}$ (26).

Finally, molecular information theory also addresses the quantitative relationship between energy dissipation during operation of a folding molecular machine and the gain in information. For a folding molecular machine operating at some physiological temperature $T_{phys}$, the bits potentially gained per operation are:

$$R_{energy}(bits/(site \cdot operation)) = \frac{-\Delta G_{folding}^{specific}}{k_B T_{phys} \cdot ln(2)} = \frac{-\Delta G_{folding}^{specific}}{\varepsilon_r} \quad (5)$$

where $\Delta G_{folding}^{specific}$ is the per site free energy of folding and $\varepsilon_r = k_B T_{phys} \cdot ln(2)$ is the actual energy dissipation required to gain one bit of information (28, 29, 41, 42). We can also define $\varepsilon_{min}$ as the minimum energy dissipation required to gain one bit of information (28, 29, 41, 42).

The efficiency of a folding molecular machine is the ratio between the minimum and the actual energy dissipated to gain one bit of information per operation $\varepsilon_{min}/\varepsilon_r$. As long as the probabilities in equation (2) refer to the same microscopic states (which is a reasonable assumption in the case of molecular machines) the increase in the information of a folding molecular machine corresponds to a decrease in entropy. In other words, for a molecular machine to make choices, its entropy must decrease. To drive this decrease, the second law of thermodynamics dictates that the entropy of the surroundings must increase more by dissipation of energy and $\varepsilon_{min} \leq \varepsilon_r$ (28, 29, 41, 42):

$$Efficiency = \frac{\varepsilon_{min}}{\varepsilon_r} \leq 1 \qquad (6)$$

Information theory further limits the information that a folding molecular machine with a certain $\Delta G_{folding}^{specific}$ can handle. When translated into molecular biology, Shannon's channel capacity theorem states that "By increasing the number of independently moving parts that can interact cooperatively to make decisions, a molecular machine can reduce the error frequency (rate of incorrect choices) to whatever arbitrarily low level is required for survival of the organism". The channel capacity theorem shows that it is possible for a protein to almost completely avoid folding into the wrong shapes, with a theoretical upper limit to the efficiency of *ln(2)* or approximately 0.693 (41). To what degree proteins approach this maximal efficiency will depend on the evolutionary pressure towards folding for a given protein and organism. Following this reasoning we can define $\varepsilon_{min}=k_B T_{sel} \cdot ln(2)$, where the effective selection temperature $T_{sel}$ is inversely proportional to the evolutionary pressure towards folding. It follows that from molecular information theory:

$$Efficiency = \frac{T_{sel}}{T_{phys}} \leq ln(2) \qquad (7)$$

The actual efficiency for the conversion of energy into information for a given folding molecular machine can be calculated from the effects of mutations on the folding process as follows. As expected from molecular information theory (32, 40), it has been repeatedly reported (47–50) that the change in folding free energy upon mutation correlates with the change in $R_{sequence}$ upon mutation, where $\Delta R_{sequence}$ can be calculated using single-site models or also take into account correlations between pairs of sites. Usually, $\Delta R_{sequence}$ is multiplied by $-k_B T_{phys} \cdot ln(2)$ to calculate a predicted change in free energy of folding (47–50). In this case, the slope of a plot of $\Delta \Delta G_{folding}^{experiment}$ versus $\Delta \Delta G_{folding}^{predicted}$ would equal $T_{sel}/T_{phys}$, that is, the efficiency for the conversion of energy into information. It follows that we can extract the efficiency for a given system from the slope of a plot of experimental versus predicted $\Delta \Delta G_{folding}$-values.

We can now define the amount of per site information effectively gained in a folding process as a function of the per site free energy of folding and the energy-to-information efficiency:

$$R_{energy}^* = R_{energy} \cdot Efficiency = \frac{-\Delta G_{folding}^{specific}}{\varepsilon_r} \cdot \frac{T_{sel}}{T_{phys}} \qquad (8)$$

Finally, molecular information theory posits that the amount of information gained by a folding molecular machine per site and operation should be the same regardless of whether it is calculated as a choice of one conformation in the pool of all polypeptide conformations, as the footprint left in protein sequences by evolution for folding, or as the amount of free energy that is effectively converted into information during folding (26, 41, 42):

$$R_{Levinthal} = R_{sequence} = R^*_{energy} \tag{9}$$

Evolution of a protein sequence for specific folding will eventually leave an information footprint of $R_{sequence}$ bits/(site operation). With each operation, the machine will gain $R_{Levinthal}$ bits/(site operation) upon folding into a specific conformation. This operation will be associated with the dissipation of $-\Delta G_{folding} = R^*_{energy} \cdot \varepsilon_r / Efficiency$ energy units per site.

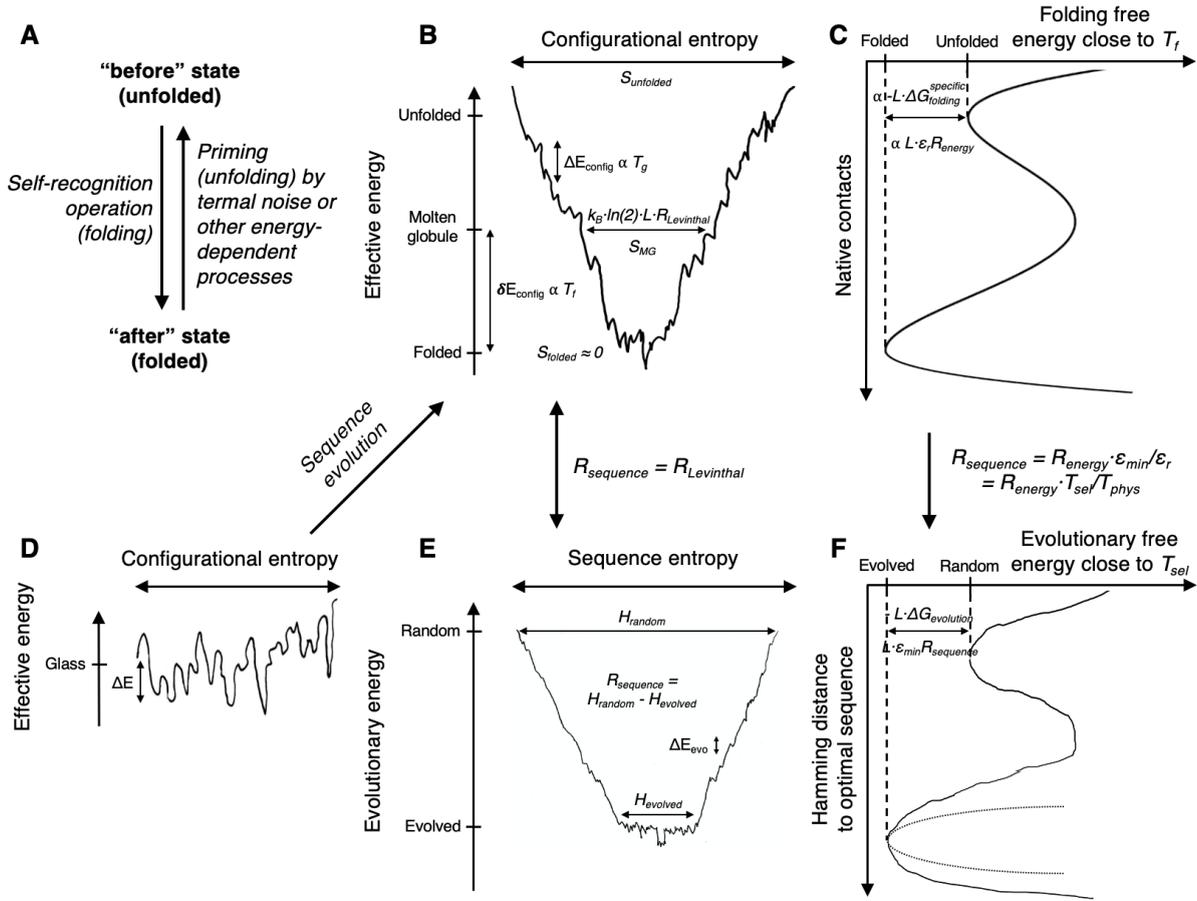

**Figure 1. Folding of a molecular machine.** (A) Operation cycle of a folding molecular machine. (B) Protein folding funnel. $S_{unfolded}$: configurational entropy of the unfolded state. $S_{MG}$: configurational entropy of the molten globule. $S_{folded}$: configurational entropy of the folded state. $\delta E_{config}$: energy gap between the molten globule and the native state. $T_f$: folding temperature, at which half the molecule is folded. $\Delta E_{config}$: landscape energy fluctuations. $T_g$: glass transition temperature. $R_{Levinthal}$: per site amount of information gained by a folding molecular machine upon finding the native configuration. $N$: number of residues in a protein. (C) Free energy profile for folding of a protein. $\Delta G_{folding}^{specific}$: per site change in free energy for folding. $N$: number of residues in a protein. $R_{energy}$: per site amount of information potentially gained by a folding molecular machine upon energy dissipation. $\varepsilon_r = k_B T_{phys} \cdot \ln(2)$: actual energy dissipation required to gain one bit of information. (D) Energy landscape for a random polypeptide. (E) Protein sequence funnel. $H_{random}$ sequence entropy of random polypeptides. $H_{evolved}$: sequence entropy of sequences evolved to fold into the target structure. $\Delta E_{evo}$: landscape energy fluctuations. (F) Hypothetical evolutionary free energy profiles for folding of a protein, with (continuous line) and without (dotted line) an evolutionary free energy barrier. $\Delta G_{evolution}$: per site change in evolutionary free energy between a random and an evolved sequence. $R_{sequence}$: per site amount of information gained by a folding molecular machine upon evolution for folding. $\varepsilon_{min} = k_B T_{sel} \cdot \ln(2)$: minimal energy dissipation required to gain one bit of information. $N$: number of residues in a protein.

## 2.2. Relationship between molecular information theory and energy landscape theory

Folding energy landscapes can be quantitatively characterized by applying the statistical mechanics of spin glasses to peptide chains (19, 51, 52). The energy landscape for folding of a protein exists in a very high dimensional space but may be represented in a simplified manner in the shape of a funnel (Figure 1B). The width of the funnel measures the configurational entropy, which decreases top to bottom as the protein approaches the folded native state. The unfolded state on top of the funnel has a large configurational entropy, while the configurational entropy of the native state at the bottom of the funnel $S_{folded}$ approaches zero. The vertical axis represents an effective energy $E_{config}$ of individual protein configurations, averaging over the solvent coordinates. According to the minimal frustration principle, these energies strongly correlate with the fraction of native contacts in each configuration.

Stabilization of the native state is allowed by the existence of an energy gap $\delta E_{config}$ with respect to the average, molten globule, compact configurations of the chain. The folding temperature $T_f$ at which half the molecule is folded is directly related to $\delta E_{config}$ and can be linked to the strength of native contacts and local structural signals. On the other hand, conformational changes between misfolded states with non-native interactions depend on the square of the landscape energy fluctuations $\Delta E_{config}$. Using a random energy model, one can define a glass temperature $T_g$ that is proportional to $\Delta E_{config}$ and describes the thermodynamics of trapping at some distance from the native structure in the configurational space. For the set of states with least structural similarity to the native state, the ratio of the folding and the glass temperature can be calculated as (22):

$$\frac{T_f}{T_g} \sim \frac{\delta E_{config}}{\Delta E_{config}} \sqrt{\frac{2k_B}{S_{unfolded}}} \qquad (10)$$

where $S_{unfolded}$ is the configurational entropy of the unfolded state and $k_B$ the Boltzmann constant. For fast and robust folding, the protein needs $\frac{T_f}{T_g} > 1$ (51), i.e. the energy landscape should be minimally frustrated.

We can relate molecular information theory to a protein folding funnel through the discussion of $R_{Levinthal}$, which measures the per site amount of information required to find the native conformation. The change in the configurational entropy of a protein upon folding $\Delta S_{folding} = S_{folded} - S_{unfolded}$ is an upper limit for $k_B \cdot ln(2) \cdot L \cdot R_{Levinthal}$ (Figure 1B), with L being the number of residues in the protein. $k_B \cdot ln(2) \cdot L \cdot R_{Levinthal}$ is likely to be significantly smaller than $\Delta S_{folding}$. For example, current protein design algorithms are able to recover most side chain conformations when given the backbone configuration of a folded protein (46, 53). From this viewpoint, some contributions to protein configurational entropy need not be included in a calculation of $R_{Levinthal}$. Under conditions favoring stability of the native state, folding effectively takes place from the molten globule and the effective change in configurational entropy upon folding is $S_{MG}-S_{folded} \approx S_{MG}$ (54). Following this reasoning, we propose that $k_B \cdot ln(2) \cdot L \cdot R_{Levinthal}$ equals the entropy of the molten globule state $S_{MG}$ (Figure 1B).

The energy and entropy in a protein folding landscape oppose each other so that at high temperature the protein is found in an ensemble of states near the top of the funnel, the unfolded state. At low temperature, an ensemble clustered around the native structures becomes thermally occupied at the bottom of the funnel. The imperfect matching of entropy and energy leads typically to a free energy barrier that separates these two ensembles of states, as depicted in Figure 1C close to $T_f$. The height of this barrier limits the folding rate and the difference in free energy between the folded and unfolded states is proportional to the protein length and the specific contribution of side chains to the free energy $\Delta G_{folding} \propto L \cdot \Delta G_{folding}^{specific}$. Equation (5) provides an additional connection between molecular information theory and energy landscape theory by relating $\Delta G_{folding}$ to $R_{energy}$ at $T_{phys}$ (Figure 1C).

Folding stability can be regarded as a physicochemical constraint on protein evolution. In the absence of such constraints, the energy landscape of random polypeptides can be described using the random energy model (55, 56). As a result of the random interactions the system behaves like a viscous liquid above $T_g$. Below this threshold, the system runs out of entropy and its kinetics exhibit glass-like behavior. Only a few paths between configurations are accessible, and several paths may lead to dead ends instead of to the native conformation. The overall shape of the energy landscape is a rugged surface (Figure 1D). As a consequence, random polypeptides do not fold in biologically relevant timescales, i.e. they do not perform the specific self-recognition operation characteristic of a folding molecular machine.

Sequence evolution of folded natural proteins from random sequences (arrow between Figure 1D and 1B) can be understood as the process leading to the appearance of a funneled energy landscape (Figure 1B). If we assume that there are no other restrictions, evolved sequences are those whose effective physical energy for the native conformation is below a certain energy threshold. The statistical mechanics equivalence between microcanonical and canonical ensemble allows one to define a selection temperature $T_{sel}$, that is not a physical temperature but a measure of the strength of selection for stability during evolution (49). For $T_{sel} \to \infty$ sequences are unconstrained by folding stability and the evolutionary landscape is frustrated. As $T_{sel}$ decreases, selection gets stronger. In particular, if $T_{sel} < T_g$ evolved sequences can fold to the native conformation (55) and the evolutionary landscape acquires an overall funnel-like shape (Figure 1E). In this case, the width of the funnel measures the sequence entropy, which decreases top to bottom as the protein takes on a more native-like sequence. The random state on top of the funnel has a large sequence entropy ($H_{random}$), while the sequence entropy of the evolved state at the bottom of the funnel ($H_{evolved}$) is still significant as observed in the sequence diversity of protein families (57). The vertical axis represents the energy $E_{evo}$ that a particular sequence has when it takes on the particular target structure. From the viewpoint of molecular information theory $R_{sequence}$, the gain of information during evolution for folding of a protein, is the difference between $H_{random}$ and $H_{evolved}$ (equation 4). Molecular information theory provides a quantitative connection between the protein folding funnel in Figure 1B and the sequence folding funnel in figure 1E through equation (9), namely that the difference in width between the top and the bottom of the two funnels is the same ($R_{sequence}=R_{Levinthal}$) (arrow between Figure 1E and 1B).

The evolutionary free energy barriers for sequence evolution are only beginning to be studied quantitatively, both from the energy landscape viewpoint (58) and from molecular information theory (35). A hypothetical transition between a random polypeptide and an evolved protein is depicted in Figure 1F close to $T_{sel}$. In some cases, there may be an evolutionary free energy barrier separating the two sequence ensembles (continuous line). However, it is also possible that sequence evolution takes place in a downhill, barrierless scenario (59, 60) (dotted line). The difference in evolutionary free energy between the random and evolved states is $\Delta G_{evolution}$. This quantity is related to $R_{sequence}$ via the scaling factor $\varepsilon_{min}=k_B T_{sel} \cdot ln(2)$ (the minimal energy dissipation required to gain one bit of information). $R_{sequence}$ is related to $R_{energy}$ via equations (5) to (7) (arrow between Figure 1F and 1C).

Finally, we consider four characteristic protein temperatures and their relationships in energy landscape theory and molecular information theory. The selection temperature $T_{sel}$ measures the strength of selection for stability during evolution. As long as $T_{sel}$ is lower than the glass temperature $T_g$, a protein sequence can evolve to fold into a globular conformation. Such a protein would remain more than 50% folded below the folding temperature $T_f$. Most natural proteins function at a physiological temperature $T_{phys}$, which is between $T_g$ and $T_f$. Altogether, we expect that $T_f > T_{phys} > T_g > T_{sel}$ so that evolved protein sequences are folded and active at the physiological temperature. According to molecular information theory, the ratio of $T_{sel}$ and $T_{phys}$ is the efficiency for conversion of free energy into information (equation 7). Energy landscape theory (49, 55) constrains the relationship between $T_f$, $T_g$ and $T_{sel}$ as follows:

$$\frac{2}{T_f T_{sel}} = \frac{1}{T_g^2} + \frac{1}{T_f^2} \tag{11}$$

In sum, we can propose a simple mapping between molecular information theory and energy landscape theory. In the next sections, we explore the connections between our current knowledge of protein folding and the quantities $R_{sequence}$, $R_{Levinthal}$ and $R^*_{energy}$.

## 2.3. Information footprint of protein folding on protein sequences: $R_{sequence}$

Here, we present and briefly discuss several estimations for the gain of information during evolution for folding of a protein, measured in bits per protein site and operation. We present calculations of $R_{sequence}$ using both aligned and unaligned protein sequences, previous results involving reduced amino acid alphabets and protein design considerations.

### 2.3.1. Information in unaligned protein sequences

A possible way to obtain a crude estimate for $R_{sequence}$ is to study databases of unaligned sequences using Kolmogorov complexity (61), k-tuplet analysis, Zipf analysis and a Chou-Fasman gambler algorithm (62). Tiana and coworkers calculate the Kolmogorov complexity of a sequence using a correlation length of one and obtain a value of 4.33 bits/site (61). However, longer correlation lengths are likely relevant for the analysis. Accordingly, k-tuplet and Zipf analysis give upper limits for the entropy of natural sequences, $H_{evolved}$, of approximately 2.5 bits/site when analyzing correlation lengths of up to four residues (62). In this case, $R_{sequence}$ is the difference between the entropy of a random amino acid mixture of twenty elements ($H_{random}$, approximately 4.32 bits/site) and the entropy of natural sequences ($H_{evolved}$). This implies that $R_{sequence}$ is at least 1.82 bits/(site operation). The Chou-Fasman gambler algorithm uses both sequence and secondary structure information to guess at the number of possible amino acids that could appropriately substitute into a sequence and yields an $H_{evolved}$ of approximately 2 bits/site. From this, $R_{sequence}$ is around 2.32 bits/(site operation). The estimations of $R_{sequence}$ presented in this section are summarized on the left side of Figure 2 and in Table 1.

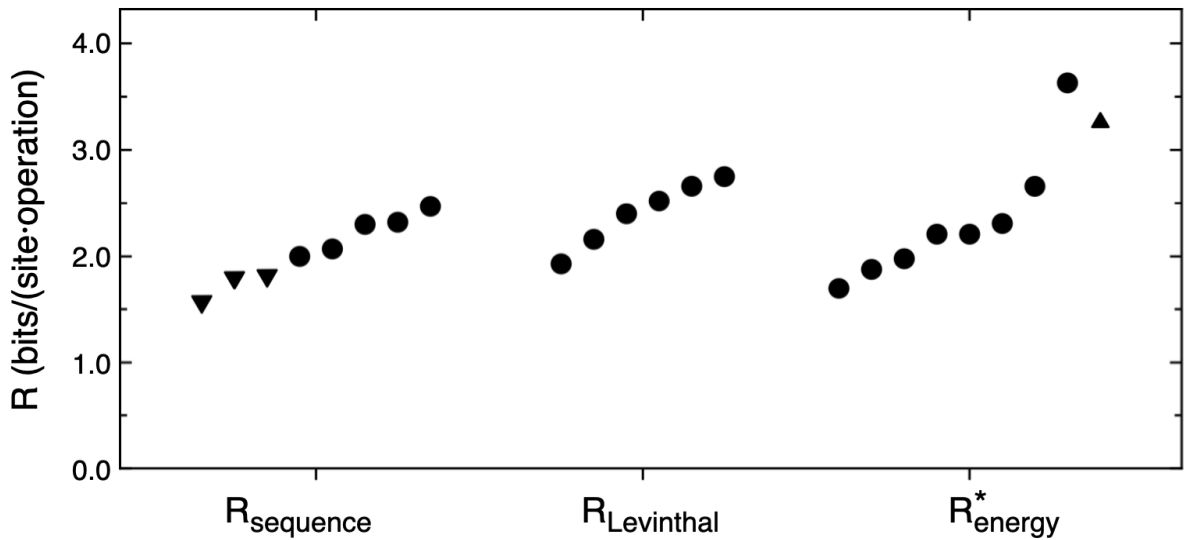

**Figure 2. Information theoretic analysis of protein folding.** Left: Information footprint of protein folding on protein sequences, as measured by $R_{sequence}$. Center: Information gain for finding the native configuration among all possible configurations, as measured by $R_{Levinthal}$. Right: Information gain associated with energy dissipation, as measured by $R^{*}_{energy}$. Circles: reported value from a method. If multiple values are reported in a source, the average is plotted. Triangles: upper limit. Inverted triangles: lower limit.

| $R_{sequence}$ calculation | $R_{sequence}$ (bits/(site operation) | $N_{evolved}$ (amino acids) | References |
|---|---|---|---|
| Unaligned sequences | >1.82, 2.32 | <5.66, 4.01 | (62) |
| Aligned sequences (single-site models) | >1.80[a], >1.57[a] | <6.03, <7.07 | This work |
| Aligned sequences (models with site correlations) | 2.47, 2.07, 2.30 | 3.79, 5.00, 4.26 | (63–65) |
| Reduced amino acid alphabets | 2.00 | 5.00 | (66–72) |
| REM model designability condition | <2.16 | >4.48 | (55, 56) |
| **Average** | **2.06±0.29** | **5.0±1.1** | |
| $R_{Levinthal}$ calculation | $R_{Levinthal}$ (bits/(site operation) | $N_{unfolded}$ (conformations) | References |
| Backbone configurational entropy | 2.52, 2.66, 1.93 | 5.73, 6.32, 3.81 | (73–75) |
| Effective secondary structure alphabet | 2.40 | 5.27 | (76) |
| Structure prediction from residue burial layers | 2.15 | 4.43 | (77–79) |
| Structure prediction from residue-residue contacts | 2.75 | 6.72 | (61) |
| **Average** | **2.40±0.31** | **5.4±1.1** | |
| $R^*_{energy}$ calculation | $R^*_{energy}$ (bits/(site operation) | Efficiency | References |
| Maximum efficiency of a molecular machine | 3.16[b] | ln(2)≈0.693 | (41) |
| Predicted *vs.* experimental changes in free energy (single-site models) | 2.58, 2.14 | 0.565, 0.47 | (47, 48) |
| Predicted *vs.* experimental changes in free energy (models with site correlations) | 1.82, 2.14, 3.53 | 0.40, 0.47, 0.77 | (49, 50, 80) |
| Using empirical estimations of $T_f/T_g$ | 1.92 | 0.42 | This work |
| Using $T_{sel}$ from the comparison between sequence-based and physical potentials | 1.65, 2.23 | 0.36, 0.49 | (49, 50) |
| **Average** | **2.25±0.59** | **0.49±0.12** [c] | |

**Table 1. Information theoretic analysis of protein folding.** Top: Information footprint of protein folding on protein sequences, as measured by $R_{sequence}$. Center: Information gain for finding the native configuration among all possible configurations, as measured by $R_{Levinthal}$. Bottom: Information gain associated with energy dissipation, as measured by $R^*_{energy}$. If multiple values are reported in a source, the average is reported. [a] upper limit. [b] lower limit. [c] Average of eight empirical estimates.

### 2.3.2. Information in aligned protein sequences

As an alternative to the analysis of sequence alignments of protein families, $R_{sequence}$ can be calculated from sequence variations within natural protein families. As a first approximation, we consider that the sequence restrictions in the different sites in a protein chain are independent (48). This is equivalent to the analysis of DNA binding sites using sequence logos (27). If we consider all amino acids that are allowed at a given site equiprobable, we can calculate $R_{sequence}$ from the size of the amino acid alphabet $N$ in the unfolded and folded states using equation (4).

In principle, $R_{sequence}$ for a sequence site can range from 0 (if the folded and unfolded states present the same alphabet size) to ≈4,39 bits (if the folded state allows for a single amino acid, considering 20 equiprobable amino acids and the gap). We can calculate the effective amino acid alphabet size in the folded state using the amino acid frequencies observed in alignments of natural protein sequences (see methods for details). Briefly, we calculated the average $N_{evolved}$ over all sites with less than 50% gaps in over 13000 protein alignments from the Pfam database. The average value of $N_{evolved}$ over all alignments is 6.0±1.8 amino acids/site (Supplementary Figure 1). On the other hand, we postulate that the unfolded state has no significant conformational or sequence restrictions and its effective alphabet size $N_{random}$ is 21. Using these figures, we can calculate a value for $R_{sequence}$ of 1.80±0.50 bits/(site operation). We may also use the observed amino acid frequencies from SwissProt, which are likely dictated by amino acid metabolic costs (81), to calculate the effective alphabet size of the unfolded state. In this case, $N_{random}$ decreases to 17.86 amino acids and $R_{sequence}$ is thus 1.57±0.50 bits/(site operation).

The above calculations ignore sequence correlations between pairs of sites, leading to an underestimation of $R_{sequence}$. Recent work by Cocco and coworkers (63) estimated the per site entropy of Hidden Markov Models of Pfam sequence families taking into account site correlations. The resulting number for $H_{evolved}$ is 1.84±0.01 bits/site, which for an alphabet of 20 amino acids amounts to an $R_{sequence}$ of 2.47±0.01 bits/(site operation). Similar recent work on three repeat protein families (64) finds an average $H_{evolved}$ of 2.32 bits/site, which for an alphabet of 21 amino acids amounts to an average $R_{sequence}$ of 2.07 bits/(site operation). Additionally, Best and coworkers (65) estimated the number of sequences compatible with 10 protein families using a statistical model based on residue-residue co-evolution. They found an average $N_{evolved}$ of 4.06 amino acids/site, which for an alphabet of 20 amino acids amounts to an $R_{sequence}$ of 2.30±0.01 bits/(site operation).

### 2.3.3. Reduced amino acid alphabets

An additional alternative for the calculation of $R_{sequence}$ comes from studies on amino acid alphabets in relation to protein folding. Such studies often ask which is the minimal alphabet size that can be used in a folding-related experiment and still attain similar results to the full alphabet of twenty amino acids. The results often reach a consensus value close to five residues: Baker and coworkers showed experimentally that an alphabet of five amino acids can largely code for a folded and functional SH3 domain (66, 67). Regarding computational experiments, an alphabet of five amino acids can also code for fast folding proteins (68), for substitution matrices and contact potentials (69) and preserves 90% of the contact mutual information (70). Further, an alphabet of five to six amino acids maximizes the information gain that links secondary structure and sequence (71), and an alphabet of six amino acids is enough for the alignment of distant sequences (72). Other computational experiments require an alphabet of around ten amino acids for protein design (82) and fold recognition (83, 84). We propose that the larger minimum alphabet sizes found in these experiments may be due to incomplete optimization of the reduced alphabets used and choose to use a reduced alphabet of 5 residues for our calculations. In this case, $N_{random}$ is 20 and $N_{evolved}$ is 5, leading to a value of $R_{sequence}$ of 2 bits/(site operation).

### 2.3.4. Designability and configurational entropy

The random energy model for the energetics of protein folding has been used to analyze the problem of evolvability, i.e., under which conditions it is feasible to find sequences that fold into a given structure. For the design or natural evolution to have a chance of success, the effective alphabet size $N_{evolved}$ must be higher than the effective number of configurations of a residue in the unfolded state, $N_{unfolded}$ (55, 56):

$$N_{evolved} > N_{unfolded} \tag{12}$$

Given that $R_{sequence} = R_{Levinthal}$ (equation (9)), we can combine equations (3) and (4):

$$R_{sequence} = log_2(N_{random}) - log_2(N_{evolved}) = log_2(N_{unfolded}) - log_2(N_{native}) = R_{Levinthal} \tag{13}$$

Approximating $N_{native}$ to 1 and combining equations (12) and (13),

$$R_{sequence} < \frac{1}{2} log_2(N_{random}) \tag{14}$$

For an amino acid alphabet size of 20, the upper limit for $R_{sequence}$ (and therefore $R_{Levinthal}$) is approximately 2.16 bits/(site operation).

### 2.3.5. Summary and discussion of $R_{sequence}$ results

Our estimations of $R_{sequence}$ from aligned and unaligned natural protein sequences, reduced alphabets and protein designability considerations converge to a relatively narrow range within the minimum value of 0 and the maximum value of ≈4.32 bits/(site operation) (Figure 2 and Table 1), with an average of 2.06±0.29 bits/(site operation). Similar to what is observed in DNA binding sequences (27, 31, 33, 36), $R_{sequence}$ is distributed unevenly along protein sequences (85), indicating that protein sites are heterogeneous in terms of their relevance towards protein folding and other biological requirements.

We can use our estimations of $R_{sequence}$ and equation (4) to calculate the size of the effective amino acid alphabet in evolved sequences $N_{evolved}$, yielding a value of 5.0±1.1 amino acids (Table 1). This is equivalent to an approximately 4-fold reduction in the size of the amino acid alphabet in natural sequences with respect to a polymer without sequence restrictions. For a polypeptide of 200 amino acids, this amounts to a $10^{120}$-fold reduction from $20^{200} \approx 10^{260}$ possible sequences to $5^{200} \approx 10^{140}$ sequences compatible with a biological function. The number of sequences compatible with a biological function is thus much less than the number of possible sequences but still much larger than the number of sequences thought to have ever existed on Earth, regardless of having been synthesized by living organisms (86) or by prebiotic reactions (87).

## 2.4. Information gain for finding the native configuration among all possible configurations: $R_{Levinthal}$

In this section, we present and briefly discuss several estimations for the decrease in the conformational uncertainty during folding of a protein, measured in bits per protein site and operation. We present calculations of $R_{Levinthal}$ for several structural features of the native state that are sufficient to infer the remaining details of the folded structure, such as backbone configurational entropy, secondary structure, residue burial, and residue-residue contacts.

### 2.4.1. Backbone configurational entropy

We consider folding as the finding of one backbone conformation among all possible conformations for a residue. The change in backbone entropy upon folding has been inferred in various ways. Molecular dynamics simulations of a native protein and a realistic unfolded state ensemble can be used to calculate a change in backbone entropy. Application of this idea to the protein ubiquitin yields a $T\Delta S_{backbone}$ of 1.0-1.1 kcal/mol per residue at 300K (73). We can convert a quantity in energy units to information units (nits) by dividing by $k_B T_{phys}$, and to bits by dividing by $ln(2)$ to account for the change from natural to base 2 logarithm. Using this conversion factor, this value of $T\Delta S_{backbone}$ of translates into an $R_{Levinthal}$ of around 2.52 bits/(site operation). Alternatively, we can analyze the values of $T\Delta S_{backbone}$ that optimize the computational prediction of folding-related quantities. The optimal value of $T\Delta S_{backbone}$ for the prediction of changes in stability upon mutation (74) is 1.1 kcal/mol per residue at 298K, which translates into an $R_{Levinthal}$ of 2.66 bits/(site operation). Similarly, the optimal value of $T\Delta S_{backbone}$ for the prediction of phi values (75) is 3.3 kcal/mol for residues in regular secondary structures and 1,2 kcal/mol for other residues. Considering an average content of regular secondary structure of 70% (76) this yields an average $T\Delta S_{backbone}$ of 2.67 kcal/mol per residue, which translates into an $R_{Levinthal}$ of 1.93 bits/(site operation).

### 2.4.2. Secondary structure alphabet

We may consider folding as the finding of one secondary structure among all possible states for each residue of the chain. As a first approximation, we consider that the conformational searches in the different sites in a protein chain are independent. We can then use equation (3) to calculate $R_{Levinthal}$ as the per site information gained upon going from an effective secondary structure alphabet of size $N_{unfolded}$ in the unfolded state to a single secondary structure for each residue in the folded state ($N_{folded} = 1$).

$R_{Levinthal}$ for an amino acid site can in principle range from 0 (if the unfolded state presents a single possible secondary structure) to $log_2(N_{unfolded})$. We can calculate the effective secondary structure alphabet size in the folded state using the secondary structure abundances observed in natural protein structures (44). Performing this calculation for the data in Table 1 of (76) yields a value of 5.29 for $N_{unfolded}$. This amounts to a value for $R_{Levinthal}$ of 2.40 bits/(site operation).

### 2.4.3. Burial layers alphabet

Alternatively, we can consider folding as the finding of one burial state among all possible states for a residue. We again consider that the conformational searches in the different sites in a protein chain are independent. We can then calculate $R_{Levinthal}$ as the per site information gained upon going from a burial alphabet of size $N_{unfolded}$ in the unfolded state to a single burial state for each residue in the folded state ($N_{folded} = 1$). From the work of Pereira de Araujo et al. (77–79) the number of burial states required to define a structure is 4 to 5. Thus, upon folding, each residue of a protein adopts 1 native burial state out of 4 to 5 possible states and $R_{Levinthal}$ is 2 to 2.32 bits/(site operation).

### 2.4.4. Residue contacts

The information contained in a residue contact map is enough to recover the structure of the native protein (88). From this viewpoint, protein folding can be understood as the process of finding the native contact map among all possible contact maps. Vendruscolo and coworkers quantified the information required to predict the structure of lysozyme and the villin headpiece subdomain to an RMSD of 4.5 Angstroms (61). Their result is an $R_{Levinthal}$ of 2.75 bits/(site operation).

### 2.4.5. Summary and discussion of $R_{Levinthal}$ results

Our estimations of $R_{Levinthal}$ from backbone configurational entropy, the effective secondary structure alphabet of proteins, the effective alphabet of burial layers and the recovery of protein structures from contact maps converge to a relatively narrow range (Figure 2 and Table 1), with an average of 2.40±0.31 bits/(site operation). We can use our estimations of $R_{Levinthal}$ and equation (3) to calculate the average number of different conformations that an amino acid can access in the unfolded state $N_{unfolded}$, yielding a value of 5.4±1.1 conformations (Table 1). This effective number is not far from Levinthal's original estimation of 3 conformations per residue (45). Accordingly, a small protein of 200 residues has approximately $10^{140}$ possible conformations if the site conformations are independent. Within the space of possible conformations, some are not foldable globular shapes. In turn, it may not be possible to find suitable sequences that fold into some of the globular structures in a reasonable time period, i.e., they are not evolvable. Additionally, it is possible that some of the evolvable globular structures have yet to appear in the biosphere. Finally, the number of natural evolvable globular structures, often called folds, lies between $10^4$ and $10^5$ (89), more than a hundred orders of magnitude smaller than the number of possible conformations.

## 2.5. Information gain associated with energy dissipation: $R^*_{energy}$

In this section, we present and briefly discuss several estimations for the gain of information associated with energy dissipation during folding of a protein, measured in bits per protein site and operation. This requires an estimation of both the per site free energy for folding and of the energy-to-informacion efficiency. We extract this information from the analysis of the effect of mutations on protein stability as follows.

### 2.5.1. What is the per residue free energy of folding?

As discussed above, molecular information theory considers the transition between a "before" non-specific state and an "after" specific state. Here, we consider folding at the single residue level. The "before" sequence-unspecific folding state would include a folded protein backbone for a generic residue, while the "after" sequence-specific folding state would also include the side chain of the wild type residue. The change in free energy for the transition between these two states could then be calculated as the change in the folding free energy upon mutation of the wild type residue to alanine, plus the free energy contribution of the side chain beta carbon to folding

$$\Delta G^{specific}_{folding} = \Delta\Delta G^{wt-Ala}_{folding} + \Delta G^{\beta\ carbon}_{folding} \tag{15}$$

where larger values of $\Delta G^{specific}_{folding}$ indicate a larger per residue free energy for folding.

The effect of mutation of a wild type residue to alanine depends strongly on the mutated site (48). In order to extract an average value, we used experimental data for eight proteins that have been subject of extensive alanine mutation experiments: Arc repressor (90, 91), BPTI (92, 93), CI2 (94–96), FBP28 WW domain (97), p53 tetramerization domain (98, 99), Pin1 WW domain (100), Protein L (101) and src SH3 domain (102). The change in folding free energy for a mutation to alanine has been measured for 77 to 94% (average 87%) of the non-glycine and non-alanine residues in these eight proteins. We estimated the change in folding free energy upon mutation for the remaining residues and the contribution of the alanine beta carbons to folding using a simple energy function that depends on the number of mutated atoms and their burial in the native state (see methods and Supplementary Table 1 for details). The average value of the per residue for each protein is then calculated for each protein. The average value of $\Delta G_{folding}^{specific}$ for the eight proteins is 1.95±0.30 kcal/(mol·site) or 4.71±0.75 bits/(site operation) in units of information according to equation (5). We can use this value and the theoretical maximum for the efficiency of conversion of energy into information (equation 7) to calculate a maximum for $R_{energy}^{*}$. The result is a theoretical upper limit for $R_{energy}^{*}$ of 3.26 bits/(site operation) (Figure 2 and Table 1).

*2.5.2. Efficiency for the conversion of energy into information in protein folding using point mutations*

Sánchez-Ruiz and coworkers used a single site model to predict the changes in folding free energy for mutations involving isoleucine and valine in *Escherichia coli* thioredoxin (47). The slope of the relationship between predicted and experimental ΔΔG*folding*-values ranged from 0.42 to 0.71 (average efficiency 0.565) depending on the sequences used for prediction. Using the average value of $\Delta G_{folding}^{specific}$ yields a $R_{energy}^{*}$ of 2.66 bits/(site operation). Similar work applying the single site model to a larger database of 1685 mutants in 44 proteins (48) yields a slope of the relationship between predicted and experimental ΔΔG*folding*-values of 0.47 (Supplementary Figure 2A). Using the average value of $\Delta G_{folding}^{specific}$ we obtain a value for $R_{energy}^{*}$ of 2.21 bits/(site operation).

Recent work predicts changes in folding free energy upon mutation with models that take into account correlations between pairs of sites. Onuchic and coworkers predicted the changes in folding free energy for mutations of a PDZ domain (49). The slope of the relationship between predicted and experimental ΔΔG*folding*-values was 0.40, which corresponds to an $R_{energy}^{*}$ of 1.88 bits/(site operation). An independent analysis of the same data with a similar model (50) yields a slope of 0.47 and an $R_{energy}^{*}$ of 2.21 bits/(site operation). A study predicting the effect of point mutations on the stability of three different ankyrin repeat proteins (80) finds a slope of 0.77 for the relationship between predicted and experimental ΔΔG*folding*-values. This value is somewhat higher than the theoretical limit of 0.693 and yields a value for $R_{energy}^{*}$ of 3.63 bits/(site operation).

*2.5.3. Selection temperature inferred from energy landscape theory*

We can combine equation (7) for the efficiency for conversion of energy to information with equation (11) for the relationship between $T_f$, $T_g$ and $T_{sel}$ to express the efficiency in terms of $T_f$, $T_{phys}$ and the $T_f/T_g$ ratio:

$$Efficiency = \frac{2T_f}{T_{phys}} \cdot \frac{1}{\frac{T_f^2}{T_g^2}+1} \qquad (16)$$

The average of several estimates for $T_f/T_g$ is approximately 2 (49), while $T_f$ is approximately 16K higher than $T_{phys}$ for *E. coli* and *T. thermophilus* proteins (103–105). Plugging these empirical values into equation (16) yields a nearly $T_{phys}$-independent value for *Efficiency* around 0.42 and a value for $R^*_{energy}$ of 1.98 bits/(site operation).

Energy landscape theory provides yet another approach to estimate the efficiency: the slope of the relationship between sequence-derived potentials and physical potentials can also be interpreted as the ratio $T_{sel}/T_f$ (49). This approach has been used to analyze eight different protein families (49), yielding an average $T_{sel}$ of 108K. This corresponds to an average ratio $T_{sel}/T_{phys}$ of 0.36 for a $T_{phys}$ of 298K and an average value for $R^*_{energy}$ of 1.70 bits/(site operation). Similar work on a different set of fourteen protein families (50) yields an average average $T_{sel}$ of 145K, which corresponds to an average $T_{sel}/T_{phys}$ of 0.49 and an average value for $R^*_{energy}$ of 2.31 bits/(site operation).

*2.5.4. Summary and discussion of $R^*_{energy}$ results*

Our estimations of the conversion of energy into information from the analysis of experimental values for the change in folding free energy upon mutation and from the comparison of sequence-derived and physical potentials range from 1.70 to 3.63 bits/(site operation) (Figure 2 and Table 1), with an average of 2.32±0.60 bits/(site operation). The range of values for $R^*_{energy}$ is higher than for $R_{sequence}$ and $R_{Levinthal}$, which may originate from a significant dependence of the results on the diversity of the sequences used in the analysis (47) and from inaccuracies on the simplified physical potentials.

The average efficiency for the conversion of energy into information in folding of the proteins analyzed in this work is 0.49±0.12 (Table 1). This value is lower than the theoretical upper limit of ≈0.693, while previous estimations for protein-DNA interactions and visual pigments closely approach this upper limit (41). The fact that protein folding dissipates more free energy than that required for coding of the native structure suggests that coding for protein function may take up to 20% of the dissipated free energy (or 0.38 kcal/(mol·site)). In agreement with this proposal, the slope of the relationship between predicted and experimental $\Delta\Delta G_{folding}$-values yields an efficiency of 63% for sequence positions mainly coding for protein structure (48) (Supplementary Figure 2B), while the slope for positions mainly coding for function yields an efficiency of 18% (Supplementary Figure 2C). We interpret that efficiency for the conversion of energy into information relevant to folding is distributed unevenly along protein sequences, similar to the uneven distribution of local frustration in relation to functional sites (106).

## 2.6. Comparison of estimates for $R_{sequence}$, $R_{Levinthal}$ and $R^*_{energy}$ and general discussion

Molecular information theory posits that the analysis of a recognition process by a molecular machine from the viewpoint of sequence restrictions, structure search and energy dissipation should lead to the same number of bits. In the case of protein-DNA recognition, the identity between $R_{sequence}$ and $R_{frequency}$ (analogous to $R_{Levinthal}$) has been shown for several molecular recognizers (26, 30, 37, 39), while the analysis of $R^*_{energy}$ points at free energy-to-information conversion efficiencies close to the theoretical maximum (41). Here, we consider protein folding as a self-recognition process by a molecular machine and calculate the per site amount of bits gained upon folding using multiple approaches for each of $R_{sequence}$, $R_{Levinthal}$ and $R^*_{energy}$. Our average estimates for $R_{sequence}$ (2.06±0.29 bits/(site operation)) and $R_{Levinthal}$ (2.40±0.31 bits/(site operation)) are in agreement with each other, supporting the information theoretic analysis of protein folding.

The identity between the information required to locate the native state in a conformational search and the information present in natural sequences can be understood as a quantitative formulation of Anfinsen's "thermodynamic hypothesis" that the information required for folding in a given environment is contained in the sequence of amino acids (11, 46, 61). Because $R_{Levinthal}$ and $N_{unfolded}$ are fixed by protein chemistry and the genetic code, $R_{sequence}$ and $N_{evolved}$ must evolve toward $R_{Levinthal}$ and $N_{unfolded}$ (35). According to a random energy model for the polypeptide energy landscape, the effective alphabet size $N_{evolved}$ must be higher than the effective number of configurations of a residue in the unfolded state, $N_{unfolded}$ (55, 56). We find that $N_{evolved}$ = 5.0±1.1 amino acids, while $N_{unfolded}$ = 5.4±1.1 conformations (Table 1). This suggests that the effective alphabet size in evolved proteins is just enough to ensure that foldable sequences can be found for the globular structures observed in nature. Folding molecular machines based on a different set of monomers would present different values of $R_{Levinthal}$ and $N_{unfolded}$ but are likely to evolve in a similar fashion as our biosphere.

Our average estimate for $R^*_{energy}$ is 2.32±0.60 bits/(site operation), in agreement with those for $R_{sequence}$ (2.06±0.29 bits/(site operation)) and $R_{Levinthal}$ (2.40±0.31 bits/(site operation)). This empirical verification of equation (9) for folding molecular machines brings together previous results from experiment and computation in different areas of protein science. Energy landscape theory contributed with an estimation of $R_{sequence}$ from structure designability considerations and estimations of $T_{sel}$ from the comparison between sequence-derived potentials and physical potentials and from the relationship between $T_f$, $T_g$ and $T_{sel}$. The statistics of protein sequences and reduced amino acid alphabets also contribute to the calculation of $R_{sequence}$. Backbone configurational entropy, the effective alphabets for protein secondary structure and burial layers and structure prediction from residue-residue contacts were also considered in the calculation of $R_{Levinthal}$. Last, sequence-based potentials and mutational studies of protein stability led to estimations of $R^*_{energy}$.

The present work brings together two conceptual frameworks. Molecular information theory can be read as a quantitative characterization of the code for protein folding and provides new connections between protein folding funnels and the corresponding energy profiles (Figure 1). Energy landscape theory may help us understand the dynamics of the gain of information during protein folding transitions, which fall mostly out of the realm of molecular information theory. We find it encouraging that it is possible to write the efficiency of a folding molecular machine in terms of the glass transition temperature and the folding transition temperature, two important parameters in energy landscape theory. It seems intriguing that the alphabet of 20 amino acids seems just large enough to fulfill both the designability/evolvability condition from equation (12) and the identity between $R_{Levinthal}$ and $R_{sequence}$ (section 2.3.4). This implies a specific value for the per site gain of information upon folding around the middle of the $[0, log_2(20)]$ interval at 2.16 bits/(site·operation), which is in agreement with the average of our estimations. In turn, this implies a minimum of value of 1.30 kcal/(mol·site) for $\Delta G^{specific}_{folding}$ if we assume the maximal theoretical efficiency for conversion of energy into information.

## 3. Concluding remarks and speculations

We propose an application of molecular information theory to protein folding. We developed here only the most basic analogy, simplifying folding of single domain proteins to a two-state transition that corresponds with the two only possible states of a molecular machine: unspecific unfolded 'before' or specific folded 'after' states. The metaphor of proteins as Machines is very common in molecular biology education and public understanding (107, 108), but these usually refer to the action of proteins once they are folded and functional. We propose that the protein folding machine is made by the system of interactions of the polypeptide chain and the solvent that self-organizes into a discrete ensemble of structural states, computing the folding problem that the primary structure poses. This must physically correspond with the collective organization of a myriad of weak interactions that cooperatively reinforce each other as they approach the native 'after' state. An elegant treatment of the initial events of protein folding from an information-processing viewpoint was conceived by Bohr and Wolynes (109). In turn, the structures pose the problem of finding which sequences are evolvable and compatible with its energetics. The average information contained in the sequences of evolved proteins ($R_{sequence}$) is very close to average information needed to specify a fold ($R_{Levinthal}$), as early envisioned by Crick (10), proposed in the thermodynamic hypothesis (11), and explained by the energy landscape theory (51). The coding of such information must be efficient enough to allow for retrieval in the noisy cellular environment. As noted before (22), such codes are expected to be non-local, degenerate and fuzzy, suitable to encode a wide range of structures with sufficient discrimination in their energy landscapes, akin to Hopfield networks (110). The relative success of different machine learning approaches to compute a tertiary structure of a natural protein given an amino acid sequence encourages us in the search for simple descriptions for these digital to analog codes (56). It would be fun if we are able to code an arbitrary problem in a sequence of amino acids and let the protein machine solve it by folding!

Molecular information theory allows for the calculation of the energy-to-information conversion efficiency of the molecular machine's operation. We estimate this efficiency at around 50%, lower than the theoretical limit of 70%, but much higher than the typical ones of human built macroscopic machines. We should bear in mind that the information expected to be coded in the sequences goes beyond the one sufficient to guide folding, as there must be specification of 'biological function' which is expected to be the ultimate level where selection pressures act. This latter informational imprint can conflict with the one needed to specify the structures, and show up as local frustration in the energy landscapes (9). It has been shown that binding, catalytic, and allosteric sites and the avoidance of aggregation may indeed conflict with the robust folding of a molecular machine (12). Presumably, making coding more effective (lowering the $T_{sel}/T_f$ ratio) would freeze the structural ensemble of the native state, impeding the visits to the excited states needed for function (111).

The values of the information content that we measured converge to an effective alphabet of ~5 residues for most natural foldable protein sequences. This is in line with previous theoretical estimates and laboratory experiments. It was suggested that as the diversity of amino acid type interactions reduces one will encounter a glass transition in sequence space (18), and finding foldable sequences would thus be biologically impeded. The fact that the current genetically coded alphabet is 20 may be related not only to the specific chemistry needed for function, but to the large entropy at the top of the sequence space funnel needed to be able to connect between different attractor folds (112). The entropy reduction going from 20 to 5 must be compensated by the effective folding free energy, which molecular information theory predicts, and we show here, is the general case. This compensation of an equilibrium thermodynamic system (protein folding) with a non-equilibrium information gathering and using system (protein evolution) (113, 114) is subtle but fundamental and is expected to occur whenever a biological phenomenon is happening. Maybe the far from equilibrium distribution of basic building blocks can be used as a biosignature in remote life-detection.

Metaphors can be very useful in scientific understanding, but we have to take the risk of making invalid analogies that may lead to incorrect mappings and not-even-wrong theories. We hope that the analogies we presented in this paper encourage the readers to examine the validity of the metaphors used in their own thinking and explore the reach of taking them seriously, but not necessarily solemnly.

## 4. Methods

### 4.1. Information in aligned protein sequences from the Pfam database

We used version 32.0 of the Pfam database for sequence alignments of protein domains (57). We considered alignments with at least 20 sequences and removed positions with more than 50% gaps to reduce uncertainties in the alignment and improve sampling. We removed alignments with a resulting length of less than 30 amino acids since they are less likely to fold into globular structures. The resulting database contains 13132 processed alignments that are 30 to 1512 amino acids long and contain 20 to 176760 sequences.

We accounted for redundancy in the aligned sequences by calculating sequence weights with the Henikoff algorithm (115). We then calculated amino acid frequencies for each position of each alignment using these sequence weights. Our calculations consider gaps as a 21st amino acid. Next, we calculated the effective number of amino acids at each position $N_{folded}$ as $2^H$, with $H$ being the sequence entropy of the position (44). Finally, we calculated the average of $N_{folded}$ for each alignment. Supplementary Figure 1 shows a histogram of the average $N_{folded}$ for all 13132 alignments. The average value is 6.03 amino acids per site (standard deviation 1.78).

### 4.2. Free energy calculations

We used eight model proteins to calculate an average value for $\Delta G_{folding}^{specific}$, Arc repressor (90, 91), BPTI (92, 93), CI2 (94–96), FBP28 WW domain (97), p53 tetramerization domain (98, 99), Pin1 WW domain (100), Protein L (101) and src SH3 domain (102). In our calculations, $\Delta G_{folding}^{specific}$ at each site of a protein is the sum of the change in folding free energy upon mutation of the wild type residue to alanine, plus the free energy contribution of the side chain beta carbon to folding (equation (15)). For the first term we used experimental data whenever available (87% of the residues). When necessary, the free energy contribution of non-mutated side chain atoms other than the beta carbon was calculated as 1.1 kcal/mol (116) times the average fraction of buried surface area of the non-mutated atoms. Residue burial was calculated using ASAView (117). If experimental data for a mutation to a non-alanine residue is available, it was taken into account and the calculation was performed for the remaining heavy atoms. The free energy contribution of the side chain beta carbon to folding was calculated as 1.1 kcal/mol (116) times the average fraction of buried surface area of non polar groups in protein structures (0.83) (118), i.e., 0.913 kcal/mol. The results of the calculation are summarized in Supplementary Table 1.

## 5. Acknowledgements

We dedicate this manuscript to Zé' Nelson Onuchic to celebrate his 60th birthday. His contributions to biological physics are only equated with the generosity he shows in sharing and nurturing deep knowledge and understanding.

This work was supported by the Consejo de Investigaciones Científicas y Técnicas (CONICET) (IES and DUF are CONICET researchers); the Agencia Nacional de Promoción Científica y Tecnológica [PICT 2016-1467 to DUF and Universidad de Buenos Aires (UBACYT 2018-20020170100540BA). Additional support from NAI and Grant Number: 80NSSC18M0093 Proposal: ENIGMA: EVOLUTION OF NANOMACHINES IN GEOSPHERES AND MICROBIAL ANCESTORS (NASA ASTROBIOLOGY INSTITUTE CYCLE 8).

# 6. References

1. Planck, Max K., *Scientific Autobiography and Other Papers* (Philosophical library, 1950).
2. P. G. Wolynes, Aperioidic crystals: Biology, Chemistry and Physics in a fugue with stretto in *AIP Conference Proceedings*, (AIP, 1988), pp. 39–65.
3. C. Taylor, B. M. Dewsbury, On the Problem and Promise of Metaphor Use in Science and Science Communication. *J Microbiol Biol Educ.* **19**, 19.1.40 (2018).
4. J. M. Smith, "The Concept of Information in Biology" in *In the Scope of Logic, Methodology and Philosophy of Science*, P. Gärdenfors, J. Woleński, K. Kijania-Placek, Eds. (Springer Netherlands, 2002), pp. 689–699.
5. Monod, Jacques, "On chance and necessity" in *Studies in the Philosophy of Biology*, (Palgrave, 1974), pp. 357–375.
6. C. Tanford, J. Reynolds, *Nature's robots: a history of proteins*, 1. issued in paperback (Oxford Univ. Press, 2003).
7. C. Adami, The use of information theory in evolutionary biology: Information theory in evolutionary biology. *Annals of the New York Academy of Sciences* **1256**, 49–65 (2012).
8. M. A. Bedau, *et al.*, Open Problems in Artificial Life. *Artificial Life* **6**, 363–376 (2000).
9. D. U. Ferreiro, E. A. Komives, P. G. Wolynes, Frustration, function and folding. *Current Opinion in Structural Biology* **48**, 68–73 (2018).
10. F. H. Crick, On protein synthesis. *Symp Soc Exp Biol* **12**, 138–163 (1958).
11. C. B. Anfinsen, Principles that Govern the Folding of Protein Chains. *Science* **181**, 223–230 (1973).
12. D. U. Ferreiro, E. A. Komives, P. G. Wolynes, Frustration in biomolecules. *Quart. Rev. Biophys.* **47**, 285–363 (2014).
13. O. Weiss, M. A. Jiménez-Montaño, H. Herzel, Information Content of Protein Sequences. *Journal of Theoretical Biology* **206**, 379–386 (2000).
14. L. Weidmann, T. Dijkstra, O. Kohlbacher, A. N. Lupas, "Minor deviations from randomness have huge repercussions on the functional structuring of sequence space" (Evolutionary Biology, 2019) https:/doi.org/10.1101/706119 (June 24, 2022).
15. P. E. Leopold, M. Montal, J. N. Onuchic, Protein folding funnels: a kinetic approach to the sequence-structure relationship. *Proc. Natl. Acad. Sci. U.S.A.* **89**, 8721–8725 (1992).
16. C. Scaletti, *et al.*, Sonification-Enhanced Lattice Model Animations for Teaching the Protein Folding Reaction. *J. Chem. Educ.* **99**, 1220–1230 (2022).
17. P. G. Wolynes, W. A. Eaton, A. R. Fersht, Chemical physics of protein folding. *Proc. Natl. Acad. Sci. U.S.A.* **109**, 17770–17771 (2012).
18. P. G. Wolynes, Evolution, energy landscapes and the paradoxes of protein folding. *Biochimie* **119**, 218–230 (2015).
19. J. D. Bryngelson, P. G. Wolynes, Spin glasses and the statistical mechanics of protein folding. *Proc. Natl. Acad. Sci. U.S.A.* **84**, 7524–7528 (1987).
20. Z. Yan, J. Wang, Funneled energy landscape unifies principles of protein binding and evolution. *Proc. Natl. Acad. Sci. U.S.A.* **117**, 27218–27223 (2020).
21. E. Bornberg-Bauer, H. S. Chan, Modeling evolutionary landscapes: Mutational stability, topology, and superfunnels in sequence space. *Proc. Natl. Acad. Sci. U.S.A.* **96**, 10689–10694 (1999).
22. R. A. Goldstein, Z. A. Luthey-Schulten, P. G. Wolynes, Optimal protein-folding codes from spin-glass theory. *Proc. Natl. Acad. Sci. U.S.A.* **89**, 4918–4922 (1992).
23. W. Lu, *et al.*, OpenAWSEM with Open3SPN2: A fast, flexible, and accessible framework for large-scale coarse-grained biomolecular simulations. *PLoS Comput Biol* **17**, e1008308 (2021).
24. J. Jumper, *et al.*, Highly accurate protein structure prediction with AlphaFold. *Nature* **596**, 583–589 (2021).
25. E. P. Wigner, G. Marx, On the future of physics. *APH N.S., Heavy Ion Physics* **1**, 87–90

(1995).
26. T. D. Schneider, G. D. Stormo, L. Gold, A. Ehrenfeucht, Information content of binding sites on nucleotide sequences. *Journal of Molecular Biology* **188**, 415–431 (1986).
27. T. D. Schneider, R. M. Stephens, Sequence logos: a new way to display consensus sequences. *Nucl Acids Res* **18**, 6097–6100 (1990).
28. T. D. Schneider, Theory of molecular machines. I. Channel capacity of molecular machines. *Journal of Theoretical Biology* **148**, 83–123 (1991).
29. T. D. Schneider, Theory of molecular machines. II. Energy dissipation from molecular machines. *Journal of Theoretical Biology* **148**, 125–137 (1991).
30. R. M. Stephens, T. D. Schneider, Features of spliceosome evolution and function inferred from an analysis of the information at human splice sites. *Journal of Molecular Biology* **228**, 1124–1136 (1992).
31. T. D. Schneider, "Reading of DNA sequence logos: Prediction of major groove binding by information theory" in *Methods in Enzymology*, (Elsevier, 1996), pp. 445–455.
32. T. D. Schneider, Information Content of Individual Genetic Sequences. *Journal of Theoretical Biology* **189**, 427–441 (1997).
33. P. Hengen, Information analysis of Fis binding sites. *Nucleic Acids Research* **25**, 4994–5002 (1997).
34. T. D. Schneider, Measuring Molecular Information. *Journal of Theoretical Biology* **201**, 87–92 (1999).
35. T. D. Schneider, Evolution of biological information. *Nucleic Acids Research* **28**, 2794–2799 (2000).
36. T. D. Schneider, Strong minor groove base conservation in sequence logos implies DNA distortion or base flipping during replication and transcription initiation. *Nucleic Acids Research* **29**, 4881–4891 (2001).
37. R. K. Shultzaberger, R. E. Bucheimer, K. E. Rudd, T. D. Schneider, Anatomy of Escherichia coli ribosome binding sites. *Journal of Molecular Biology* **313**, 215–228 (2001).
38. T. D. Schneider, Claude Shannon: Biologist [information theory used in biology]. *IEEE Eng. Med. Biol. Mag.* **25**, 30–33 (2006).
39. R. K. Shultzaberger, Z. Chen, K. A. Lewis, T. D. Schneider, Anatomy of Escherichia coli σ 70 promoters. *Nucleic Acids Research* **35**, 771–788 (2007).
40. R. K. Shultzaberger, *et al.*, Correlation between binding rate constants and individual information of E. coli Fis binding sites. *Nucleic Acids Research* **35**, 5275–5283 (2007).
41. T. D. Schneider, 70% efficiency of bistate molecular machines explained by information theory, high dimensional geometry and evolutionary convergence. *Nucleic Acids Research* **38**, 5995–6006 (2010).
42. T. D. Schneider, A brief review of molecular information theory. *Nano Communication Networks* **1**, 173–180 (2010).
43. T. D. Schneider, V. Jejjala, Restriction enzymes use a 24 dimensional coding space to recognize 6 base long DNA sequences. *PLoS ONE* **14**, e0222419 (2019).
44. Shannon, Claude A., A mathematical theory of communication. *Bell Syst Tech Journal* **27**, 379–423 (1948).
45. Levinthal, Cyrus, How to fold graciously. *Mossbauer spectroscopy in biological systems* **67**, 22–24 (1969).
46. J. M. Brisendine, R. L. Koder, Fast, cheap and out of control — Insights into thermodynamic and informatic constraints on natural protein sequences from de novo protein design. *Biochimica et Biophysica Acta (BBA) - Bioenergetics* **1857**, 485–492 (2016).
47. R. Godoy-Ruiz, R. Perez-Jimenez, B. Ibarra-Molero, J. M. Sanchez-Ruiz, A Stability Pattern of Protein Hydrophobic Mutations that Reflects Evolutionary Structural Optimization. *Biophysical Journal* **89**, 3320–3331 (2005).
48. I. E. Sánchez, J. Tejero, C. Gómez-Moreno, M. Medina, L. Serrano, Point Mutations in Protein Globular Domains: Contributions from Function, Stability and Misfolding. *Journal of Molecular Biology* **363**, 422–432 (2006).

49. F. Morcos, N. P. Schafer, R. R. Cheng, J. N. Onuchic, P. G. Wolynes, Coevolutionary information, protein folding landscapes, and the thermodynamics of natural selection. *Proc. Natl. Acad. Sci. U.S.A.* **111**, 12408–12413 (2014).
50. S. Miyazawa, Selection originating from protein stability/foldability: Relationships between protein folding free energy, sequence ensemble, and fitness. *Journal of Theoretical Biology* **433**, 21–38 (2017).
51. J. N. Onuchic, Z. Luthey-Schulten, P. G. Wolynes, THEORY OF PROTEIN FOLDING: The Energy Landscape Perspective. *Annu. Rev. Phys. Chem.* **48**, 545–600 (1997).
52. J. Nelson Onuchic, H. Nymeyer, A. E. García, J. Chahine, N. D. Socci, "The energy landscape theory of protein folding: Insights into folding mechanisms and scenarios" in *Advances in Protein Chemistry*, (Elsevier, 2000), pp. 87–152.
53. L. L. Looger, H. W. Hellinga, Generalized dead-end elimination algorithms make large-scale protein side-chain structure prediction tractable: implications for protein design and structural genomics. *Journal of Molecular Biology* **307**, 429–445 (2001).
54. Z. Luthey-Schulten, B. E. Ramirez, P. G. Wolynes, Helix-Coil, Liquid Crystal, and Spin Glass Transitions of a Collapsed Heteropolymer. *J. Phys. Chem.* **99**, 2177–2185 (1995).
55. V. S. Pande, A. Y. Grosberg, T. Tanaka, Statistical mechanics of simple models of protein folding and design. *Biophysical Journal* **73**, 3192–3210 (1997).
56. G. Magi Meconi, I. Sasselli, V. Bianco, J. Onuchic, I. Coluzza, Key aspects of the past 30 Years of protein design. *Rep. Prog. Phys.* (2022) https:/doi.org/10.1088/1361-6633/ac78ef (June 23, 2022).
57. J. Mistry, *et al.*, Pfam: The protein families database in 2021. *Nucleic Acids Research* **49**, D412–D419 (2021).
58. A. R. Kinjo, Cooperative "folding transition" in the sequence space facilitates function-driven evolution of protein families. *Journal of Theoretical Biology* **443**, 18–27 (2018).
59. J. D. Bryngelson, J. N. Onuchic, N. D. Socci, P. G. Wolynes, Funnels, pathways, and the energy landscape of protein folding: A synthesis. *Proteins* **21**, 167–195 (1995).
60. V. Muñoz, L. A. Campos, M. Sadqi, Limited cooperativity in protein folding. *Current Opinion in Structural Biology* **36**, 58–66 (2016).
61. A. Possenti, M. Vendruscolo, C. Camilloni, G. Tiana, A method for partitioning the information contained in a protein sequence between its structure and function. *Proteins* **86**, 956–964 (2018).
62. B. J. Strait, T. G. Dewey, The Shannon information entropy of protein sequences. *Biophysical Journal* **71**, 148–155 (1996).
63. J. P. Barton, A. K. Chakraborty, S. Cocco, H. Jacquin, R. Monasson, On the Entropy of Protein Families. *J Stat Phys* **162**, 1267–1293 (2016).
64. J. Marchi, *et al.*, Size and structure of the sequence space of repeat proteins. *PLoS Comput Biol* **15**, e1007282 (2019).
65. P. Tian, R. B. Best, How Many Protein Sequences Fold to a Given Structure? A Coevolutionary Analysis. *Biophysical Journal* **113**, 1719–1730 (2017).
66. D. S. Riddle, *et al.*, Functional rapidly folding proteins from simplified amino acid sequences. *Nat Struct Mol Biol* **4**, 805–809 (1997).
67. P. G. Wolynes, As simple as can be? *Nat Struct Mol Biol* **4**, 871–874 (1997).
68. W. Wang, J. Wang, A computational approach to simplifying the protein folding alphabet. *Nat. Struct Biol.* **6**, 1033–1038 (1999).
69. F. Melo, M. A. Marti-Renom, Accuracy of sequence alignment and fold assessment using reduced amino acid alphabets. *Proteins* **63**, 986–995 (2006).
70. A. D. Solis, Amino acid alphabet reduction preserves fold information contained in contact interactions in proteins: Amino Acid Alphabet Reduction. *Proteins* **83**, 2198–2216 (2015).
71. A. D. Solis, S. Rackovsky, Optimized representations and maximal information in proteins. *Proteins* **38**, 149–164 (2000).
72. J. Li, W. Wang, Grouping of amino acids and recognition of protein structurally

conserved regions by reduced alphabets of amino acids. *SCI CHINA SER C* **50**, 392–402 (2007).
73. M. C. Baxa, E. J. Haddadian, J. M. Jumper, K. F. Freed, T. R. Sosnick, Loss of conformational entropy in protein folding calculated using realistic ensembles and its implications for NMR-based calculations. *Proc. Natl. Acad. Sci. U.S.A.* **111**, 15396–15401 (2014).
74. R. Guerois, J. E. Nielsen, L. Serrano, Predicting Changes in the Stability of Proteins and Protein Complexes: A Study of More Than 1000 Mutations. *Journal of Molecular Biology* **320**, 369–387 (2002).
75. V. Muñoz, W. A. Eaton, A simple model for calculating the kinetics of protein folding from three-dimensional structures. *Proc. Natl. Acad. Sci. U.S.A.* **96**, 11311–11316 (1999).
76. C. A. F. Andersen, A. G. Palmer, S. Brunak, B. Rost, Continuum Secondary Structure Captures Protein Flexibility. *Structure* **10**, 175–184 (2002).
77. J. R. Rocha, M. G. van der Linden, D. C. Ferreira, P. H. Azevêdo, A. F. Pereira de Araújo, Information-theoretic analysis and prediction of protein atomic burials: on the search for an informational intermediate between sequence and structure. *Bioinformatics* **28**, 2755–2762 (2012).
78. M. G. van der Linden, D. C. Ferreira, L. C. de Oliveira, J. N. Onuchic, A. F. Pereira de Araújo, Ab initio protein folding simulations using atomic burials as informational intermediates between sequence and structure: Folding Simulations with Atomic Burial Predictions. *Proteins* **82**, 1186–1199 (2014).
79. D. C. Ferreira, M. G. van der Linden, L. C. de Oliveira, J. N. Onuchic, A. F. Pereira de Araújo, Information and redundancy in the burial folding code of globular proteins within a wide range of shapes and sizes: Information and Redundancy in the Burial Folding Code. *Proteins* **84**, 515–531 (2016).
80. E. A. Galpern, J. Marchi, T. Mora, A. M. Walczak, D. U. Ferreiro, From evolution to folding of repeat proteins (2022) https:/doi.org/10.48550/ARXIV.2202.12223 (June 13, 2022).
81. T. Krick, *et al.*, Amino Acid Metabolism Conflicts with Protein Diversity. *Molecular Biology and Evolution* **31**, 2905–2912 (2014).
82. K. Fan, W. Wang, What is the Minimum Number of Letters Required to Fold a Protein? *Journal of Molecular Biology* **328**, 921–926 (2003).
83. L. R. Murphy, A. Wallqvist, R. M. Levy, Simplified amino acid alphabets for protein fold recognition and implications for folding. *Protein Engineering, Design and Selection* **13**, 149–152 (2000).
84. T. Li, K. Fan, J. Wang, W. Wang, Reduction of protein sequence complexity by residue grouping. *Protein Engineering Design and Selection* **16**, 323–330 (2003).
85. Å. Pérez-Bercoff, J. Koch, T. R. Bürglin, LogoBar: bar graph visualization of protein logos with gaps. *Bioinformatics* **22**, 112–114 (2006).
86. D. T. F. Dryden, A. R. Thomson, J. H. White, How much of protein sequence space has been explored by life on Earth? *J. R. Soc. Interface.* **5**, 953–956 (2008).
87. H. P. Yockey, Self organization origin of life scenarios and information theory. *Journal of Theoretical Biology* **91**, 13–31 (1981).
88. M. Vendruscolo, E. Kussell, E. Domany, Recovery of protein structure from contact maps. *Folding and Design* **2**, 295–306 (1997).
89. M. J. Sippl, Fold space unlimited. *Current Opinion in Structural Biology* **19**, 312–320 (2009).
90. M. E. Milla, B. M. Brown, R. T. Sauer, Protein stability effects of a complete set of alanine substitutions in Arc repressor. *Nat Struct Mol Biol* **1**, 518–523 (1994).
91. M. E. Milla, R. T. Sauer, Critical side-chain interactions at a subunit interface in the Arc repressor dimer. *Biochemistry* **34**, 3344–3351 (1995).
92. K.-S. Kim, *et al.*, Crevice-forming mutants of bovine pancreatic trypsin inhibitor: Stability changes and new hydrophobic surface: BPTI mutants, crevice mutants. *Protein Science* **2**, 588–596 (1993).

93. M.-H. Yu, J. S. Weissman, P. S. Kim, Contribution of individual side-chains to the stability of BPTI examined by alanine-scanning mutagenesis. *Journal of Molecular Biology* **249**, 388–397 (1995).
94. L. S. Itzhaki, D. E. Otzen, A. R. Fersht, The Structure of the Transition State for Folding of Chymotrypsin Inhibitor 2 Analysed by Protein Engineering Methods: Evidence for a Nucleation-condensation Mechanism for Protein Folding. *Journal of Molecular Biology* **254**, 260–288 (1995).
95. D. E. Otzen, A. R. Fersht, Folding of Circular and Permuted Chymotrypsin Inhibitor 2: Retention of the Folding Nucleus. *Biochemistry* **37**, 8139–8146 (1998).
96. D. E. Otzen, A. R. Fersht, Analysis of protein–protein interactions by mutagenesis: direct versus indirect effects. *Protein Engineering, Design and Selection* **12**, 41–45 (1999).
97. M. Petrovich, A. L. Jonsson, N. Ferguson, V. Daggett, A. R. Fersht, Φ-Analysis at the Experimental Limits: Mechanism of β-Hairpin Formation. *Journal of Molecular Biology* **360**, 865–881 (2006).
98. M. G. Mateu, M. M. Sánchez Del Pino, Fersht, Alan R., Mechanism of folding and assembly of a small tetrameric protein domain from tumor suppressor p53 M G Mateu 1, M M Sánchez Del Pino, A R Fersht. *Nat. Struct Biol.* **6**, 191–198 (1999).
99. M. G. Mateu, Fersht, Alan R., Nine hydrophobic side chains are key determinants of the thermodynamic stability and oligomerization status of tumour suppressor p53 tetramerization domain. *The EMBO Journal* **17**, 2748–2758 (1998).
100. M. Jäger, M. Dendle, J. W. Kelly, Sequence determinants of thermodynamic stability in a WW domain-An all-β-sheet protein. *Protein Science* **18**, 1806–1813 (2009).
101. D. E. Kim, C. Fisher, D. Baker, A breakdown of symmetry in the folding transition state of protein L. *Journal of Molecular Biology* **298**, 971–984 (2000).
102. V. P. Grantcharova, D. S. Riddle, J. V. Santiago, D. Baker, Important role of hydrogen bonds in the structurally polarized transition state for folding of the src SH3 domain. *Nat Struct Mol Biol* **5**, 714–720 (1998).
103. P. Leuenberger, *et al.*, Cell-wide analysis of protein thermal unfolding reveals determinants of thermostability. *Science* **355**, eaai7825 (2017).
104. A. Mateus, *et al.*, Thermal proteome profiling in bacteria: probing protein state *in vivo*. *Mol Syst Biol* **14** (2018).
105. R. Nikam, A. Kulandaisamy, K. Harini, D. Sharma, M. M. Gromiha, ProThermDB: thermodynamic database for proteins and mutants revisited after 15 years. *Nucleic Acids Research* **49**, D420–D424 (2021).
106. D. U. Ferreiro, J. A. Hegler, E. A. Komives, P. G. Wolynes, Localizing frustration in native proteins and protein assemblies. *Proc. Natl. Acad. Sci. U.S.A.* **104**, 19819–19824 (2007).
107. B. Alberts, The Cell as a Collection of Protein Machines: Preparing the Next Generation of Molecular Biologists. *Cell* **92**, 291–294 (1998).
108. D. S. Goodsell, *The machinery of life*, 2nd ed., corrected (Copernicus Books, 2009).
109. H. G. Bohr, P. G. Wolynes, Initial events of protein folding from an information-processing viewpoint. *Phys. Rev. A* **46**, 5242–5248 (1992).
110. J. J. Hopfield, Neural networks and physical systems with emergent collective computational abilities. *Proc. Natl. Acad. Sci. U.S.A.* **79**, 2554–2558 (1982).
111. H. Frauenfelder, S. G. Sligar, P. G. Wolynes, The Energy Landscapes and Motions of Proteins. *Science* **254**, 1598–1603 (1991).
112. M. J. Denton, C. J. Marshall, M. Legge, The Protein Folds as Platonic Forms: New Support for the Pre-Darwinian Conception of Evolution by Natural Law. *Journal of Theoretical Biology* **219**, 325–342 (2002).
113. W. H. Zurek, "Algorithmic Information Content, Church — Turing Thesis, Physical Entropy, and Maxwell's Demon" in *Information Dynamics*, NATO ASI Series., H. Atmanspacher, H. Scheingraber, Eds. (Springer US, 1991), pp. 245–259.
114. A. Balbín, E. Andrade, Protein Folding and Evolution are Driven by the Maxwell Demon Activity of Proteins. *Acta Biotheor* **52**, 173–200 (2004).

115. S. Henikoff, J. G. Henikoff, Position-based sequence weights. *Journal of Molecular Biology* **243**, 574–578 (1994).
116. C. N. Pace, *et al.*, Contribution of Hydrophobic Interactions to Protein Stability. *Journal of Molecular Biology* **408**, 514–528 (2011).
117. S. Ahmad, M. Gromiha, H. Fawareh, A. Sarai, ASAView: Database and tool for solvent accessibility representation in proteins. *BMC Bioinformatics* **5**, 51 (2004).
118. C. N. Pace, Polar Group Burial Contributes More to Protein Stability than Nonpolar Group Burial. *Biochemistry* **40**, 310–313 (2001).

**For table of contents only:**

Folding molecular machine 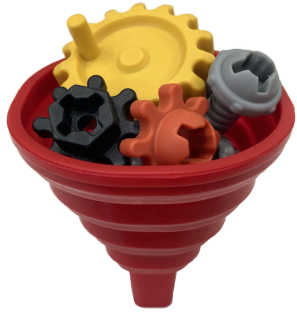

**Supplementary material.**

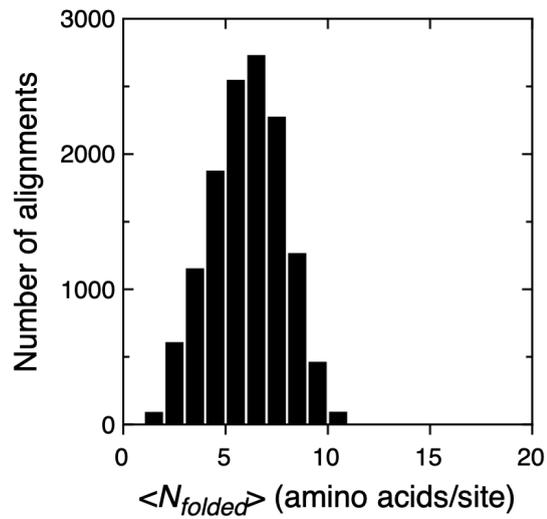

**Supplementary Figure 1.** Histogram of average $N_{folded}$ for 13132 Pfam alignments. The average value is 6.03 amino acids per site (standard deviation 1.78).

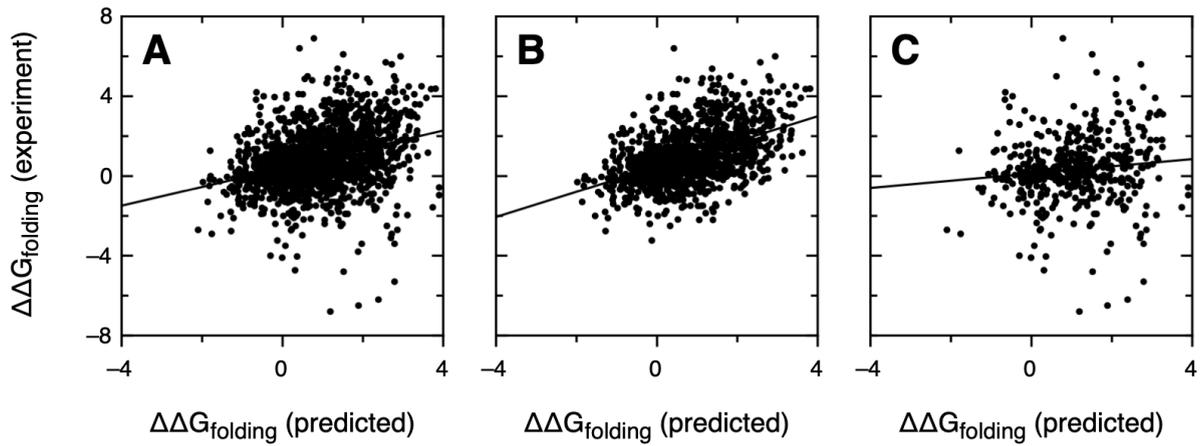

**Supplementary Figure 2.** Relationship between predicted and experimental $\Delta\Delta G_{folding}$-values using a single site model (48). The lines are linear fits to the data. (A) All protein positions (1685 point mutants). Slope is 0.47±0.03, intercept 0.39±0.05. (B) Positions mainly coding for folding (1228 point mutants). Slope is 0.63±0.03, intercept 0.47±0.05. (C) Positions mainly coding for function (457 point mutants). Slope is 0.18±0.07, intercept 0.13±0.11.

**Supplementary Table 1. Calculation of $\Delta G_{folding}^{specific}$.** See methods for details of the calculations. [a] Does not include alanine and glycine residues. Includes residues not mutated to alanine as indicated. [b] Protein is a homodimer. [c] Protein is a homotetramer. NVal stands for norvaline. [d] Site average value of $\Delta G_{folding}^{specific} = \Delta\Delta G_{folding}^{wt-Ala}(experiment) + \Delta\Delta G_{folding}^{wt-Ala}(calculated) + \Delta G_{folding}^{\beta\, carbon}(calculated)$. [e] Calculated as $\Delta G_{folding}^{specific} \div (k_B T \cdot ln(2))$. This is the maximum amount of bits that can potentially be gained per site and folding operation (equation (5)).

| Protein | Length (amino acids) | $\Sigma \Delta\Delta G_{folding}^{wt-Ala}$ experiment (kcal/mol) | Non-mutated amino acids [a] | $\Sigma \Delta\Delta G_{folding}^{wt-Ala}$ calculated (kcal/mol) | $\Sigma \Delta G_{folding}^{\beta\, carbon}$ calculated (kcal/mol) | $\Delta G_{folding}^{specific}$ calculated (kcal/(mol·site)) [d] | T (K) | $R_{energy}$ (bits/(site· operation)) [e] | References |
|---|---|---|---|---|---|---|---|---|---|
| Pin1 WW | 34 | 18.2 | W11F Y24F N26 P37 | 14.8 | 29.2 | 1.83 | 313 | 4.21 | (100) |
| FBP28 WW | 37 | 16.2 | T3 W8F K12 Y20F N22 E31T K32 P33 Q34 K37 | 22.7 | 32.0 | 1.92 | 283 | 4.88 | (97) |
| src SH3 | 57 | 34.6 | E17 E21 T38 E39 D41 | 2.8 | 48.4 | 1.51 | 295 | 3.68 | (102) |
| BPTI | 58 | 46 | C5 C14 V30 F33 C38 N43 C55 | 14.5 | 47.5 | 1.86 | 312 | 4.30 | (92, 93) |
| Protein L | 61 | 67.4 | Q18 E27 D43 Y47 D53 F62 | 10.9 | 51.1 | 2.12 | 295 | 5.19 | (101) |
| CI2 | 65 | 63.1 | N0 L1 E4 W5 V9 V13 Q28 V31 M40 I44 K53 L54 D55 E59 R62 | 27.9 | 56.6 | 2.27 | 298 | 5.50 | (94–96) |
| Arc | 102 | 62.6 | V22 E36 F45 [b] | 25.3 | 85.8 | 1.70 | 298 | 4.12 | (90, 91) |
| p53tet | 120 | 99.8 | Y327 L330NVal I332V F341L L344NVal L348NVal Q354 [c] | 82.3 | 105.9 | 2.40 | 298 | 5.81 | (98, 99) |
| Average | | 67±30 | | | | 1.95±0.30 | | 4.71±0.75 | |